\documentclass[letterpaper]{jpconf}
\usepackage{graphicx}
\begin{document}
\title{Principles of Einstein--Finsler Gravity and Cosmology}

\author{Sergiu I. Vacaru}

\address{Department of Science, University Al.I.  Cuza at Ia\c si (UAIC),\newline   54 Lascar Catargi str., Ia\c si, Romania, 700107}

\ead{sergiu.vacaru@uaic.ro, Sergiu.Vacaru@gmail.com}

\begin{abstract}
We analyze the  foundations of Finsler gravity theories
with metric compatible connections constructed on nonholonomic  tangent bundles, or (pseudo) Riemannian manifolds. There are considered "minimal" modifications of  Einstein gravity
(including theories with violation of local Lorentz invariance) and shown how the general relativity theory and   generalizations can be equivalently re--formulated in Finsler like variables.  We focus on Finsler branes solutions and perspectives in modern cosmology.
\end{abstract}

\section{Introduction}

The bulk of classical gravitational effects are described experimentally in the framework of General Relativity (GR). Recently, it was proposed that certain exceptions in theoretical cosmology may be related to the dark matter and dark energy problems and some approaches were formulated for Finlser gravity models.  It is also expected that small corrections with violations of equivalence principle and local Lorentz invariance (in brief, LV) have to be considered in low energy limits within general approaches to quantum gravity (QG).\footnote{We cite a series of works  and  reviews of results  on Finsler geometry methods and modifications of gravity and related string/brane gravity models and  QG phenomenology \cite{vstr2,perlick,vsgg,ijgmmp}, noncommutative geometry and deformation/two connection quantization \cite{vncbh}, nonholonomic Ricci flows and cosmology  \cite{vriccisp,vprincipl}.}

There are two general classes of Finsler type gravity
theories:\ The first one originates
from E. Cartan works on Finsler geometry \cite{cartan}, see
further geometrical developments and applications in \cite%
{vstr2,ijgmmp,vncbh,vprincipl,ma}. Such most related to ''standard physics'' constructions were elaborated for the metric compatible Cartan and canonical distinguished
connections (in brief, d--connections). This class of  Einstein--Finsler gravity (EFG) models can be elaborated following geometric and physical principles which are very similar to those in GR but with  Cartan--Finsler type connections used instead of the Levi--Civita connection. In the second class of theories, the Finsler connections (for instance, the Berwald/ Chern d--connections) are not metric compatible \cite{bcs,ma,vsgg}. This results in non--standard physical theories with ambiguities  in definition of  Dirac operators,  quantum/noncommutative generalizations of gravity models etc  \cite{vprincipl}.

The goals, and structure, of this paper, summarizing certain our recent results \cite{ijgmmp,vncbh,vprincipl}, are to  study Finsler modifications of GR derived from MDR of classical gravity and QG theories with LV and  formulate certain fundamental principles and modifications of axioms of GR including the EFG theories (section \ref{s2}) and speculate on Finsler/--brane models with acceleration of Universe (section \ref{s3}).
 Conclusions  are presented in section \ref{s4}.

\section{Principles  of Einstein--Finsler Gravity} \label{s2}
\subsection{Motivations and fundamental geometric/physical  objects}
  We consider four important motivations to study Finsler modifications  of gravity theories and formulations of GR in almost K\"{a}hler -- Finsler variables \cite{perlick,vprincipl}:
 \begin{enumerate}
\item  For deformations of Minkovski spacetimes resulting in nonlinear energy--momentum relations,
 $E^{2}=p^{2}c^{2}+m_{0}^{2}c^{4}+\varphi (E,p),$
 where $\ E\sim \frac{\partial }{\partial t},p_{\widehat{i}}\sim \frac{%
\partial }{\partial x^{\widehat{i}}},$   $\omega \sim %
\frac{\partial}{\partial t}$  $k_{\widehat{i}}\sim \frac{\partial}{\partial
x^{\widehat{i}}},$\
 $(x^{1}=ct,x^{2},x^{3},x^{4}); \widehat{i},\widehat{j}...=2,3,4;$ (for instance, $i,j,...=(1,\widehat{i}),(1,\widehat{j})$),
 we obtain modified dispersion relations (MDR)
 $\omega ^{2}=c^{2}[g_{\widehat{i}\widehat{j}}k^{\widehat{i}}k^{\widehat{%
j}}] ^{2}(1- {q_{\widehat{i}_{1}\widehat{i}_{2}...%
\widehat{i}_{2r}}y^{\widehat{i}_{1}}...y^{\widehat{i}_{2r}}} / r {[ g_{%
\widehat{i}\widehat{j}}k^{\widehat{i}}k^{\widehat{j}}] ^{2r}}) $. To such MDR, we can
associate  a
fundamental Finsler function $%
F(x^{i},\beta y^{j})=\beta F(x^{i},y^{j}), \beta >0,$ in tangent space $TV$, and a nondegenerated Hessian,  $\ ^{F}g_{ij}(x^{i},y^{j})=\frac{1}{2}\frac{\partial F^{2}}{\partial y^{i}\partial y^{j}},$ when
 \begin{equation}ds^{2} =F^{2}
 \approx -(cdt)^{2}+g_{\widehat{i}\widehat{j}}(x^{k})y^{\widehat{i}}y^{%
\widehat{j}} [1+\frac{1}{r}{q_{\widehat{i}_{1}\widehat{i}_{2}...%
\widehat{i}_{2r}}(x^{k})y^{\widehat{i}_{1}}...y^{\widehat{i}_{2r}}}/{(
g_{\widehat{i}\widehat{j}}(x^{k})y^{\widehat{i}}y^{\widehat{j}}) ^{r}}] +O(q^{2}). \label{finslgf}
\end{equation}
Various scenarios  in modern cosmology with dark  energy/matter, QG phenomenology with LV, corrections from string/brane  gravity, noncommutative gravity models etc are aimed to solve some important theoretical problems and  explain existing  experimental data \cite{perlick,vstr2,vprincipl}.

\item The Einstein equations in GR, and modifications, can be integrated "almost in general" forms, for arbitrary finite dimension, in terms of Finsler like connections; imposing   nonholonomic constraints, we generate solutions  for  the Levi--Civita connection \cite{ijgmmp,vsgg}.

\item Rewriting GR, equivalently,  in almost K\"{a}ehler -- Finsler variables, the theory can be quantized following methods of deformation/ geometric and A--brane quantization when a "two--connection" geometric renormalization procedure can be performed  \cite{vncbh}.
\item Finsler metrics and connections are induced via  nonholonomic and/or  noncommutative Ricci flows of Einstein spacetimes and generalizations \cite{vriccisp}.
 \end{enumerate}

{\it{ \bf Claim:}}  A (pseudo) Finsler fundamental  function $F^2$ (\ref{finslgf}) and its Hessian ("vertical" metric  in $TV$) $\ ^{F}g_{ij}(x^k,y^a)$ do not  state completely a geometric model.  A Finsler geometry/ gravity $(F: \mathbf{N,g,D})$  is defined by  three fundamental geometric objects  generated by any $F(x,y):$
 \begin{itemize}
  \item
 a nonlinear connection (N--connection) structure (i.e. a nonholonomic distributions with conventional h- / v--splitting $\ ^F\mathbf{N}=\{N^a_i(x,y)\}: TTV=hTV \oplus vTV$)  determined canonically by the Euler--Lagrange equations for  $L=F^2$ treated equivalently as semi--spray equations (in such a case, $\mathbf{N}=\ ^c\mathbf{N}$);\footnote{ there are  N--elongated (co) frames (respectively, partial derivatives and differential operators), $\mathbf{e}_{\nu } = ( \mathbf{e}_{i}=\partial _{i}-N_{i}^{a}\partial
_{a},e_{a}=\partial _{a})$ and $\mathbf{e}^{\mu}=(e^j,\mathbf{e}^b=dy^b+N_{j}^{b}dy^b),$ for h- and v--coordinates $u^\alpha =(x^i,y^a); i,j,...=1,2,...n$ and $a,b,...=n+1,n+2,...n+n$ (on local formulas see details in \cite{ijgmmp,vprincipl});}
  \item
     a distinguished metric (d--metric), $\ ^F\mathbf{g}=h\mathbf{g}\oplus  v\mathbf{g};$ up to frame transforms, it can be parametrized in N--adapted form, $\ ^{F}\mathbf{g} = \{\ ^{F}\mathbf{g}_{\alpha \beta} \} =\ ^{F}g_{ij}(u) dx^{i}\otimes dx^{j}+
\ ^{F}g_{ab}(u)  \ ^{c}\mathbf{e}^{a}\otimes
\ ^{c}\mathbf{e}^{b};$
  \item a distinguished connection  (d--connection;  a  N--adapted linear connection preserving $\mathbf{N}$ under  parallelism),  $\ ^F\mathbf{D}=(hD,vD),$ which also  determined  by  $\ ^{F}g_{ij}$ and $N^a_i$.
    \end{itemize}

A Finsler geometry can be  modelled  on a tangent bundle $\mathbf{V}=TV$ to a manifold $V$, or on a {\sf nonholonomic manifold} $\mathbf{V}=(\mathcal{V}, \mathcal{D}),$ where $\mathcal{D}$ is a nonholonomic distribution, for instance, defining a N--connection structure \cite{vsgg,vprincipl,ijgmmp}. If  $\mathcal{V}$ is a (pseudo) Riemannian spacetime with a conventional  nonholonomic $2+2$ splitting, we can introduce Finsler variables  in GR.\footnote{For any given (Finsler) d--connection  $\mathbf{D}=\{\mathbf{\Gamma }_{\ \beta
\gamma }^{\alpha }\},$  there are computed in standard forms  (see details in \cite{ma,bcs,vsgg,ijgmmp,vprincipl}) the  N-- adapted coefficients of physically important tensors:  torsion, $\mathbf{T}_{\
\beta \gamma }^{\alpha }=\{T_{\ jk}^{i},T_{\ ja}^{i},T_{\ ji}^{a},T_{\
bi}^{a},T_{\ bc}^{a}\}$,  curvature, $\mathbf{\mathbf{R}}_{\ \ \beta \gamma \delta
}^{\alpha }=\{R_{\ hjk}^{i},R_{\ bjk}^{a},R_{\ jka}^{i},R_{\
bka}^{c},R_{\ jbc}^{i},R_{\ bcd}^{a}\},$ Ricci tensor,
 $\mathbf{R}_{\alpha \beta }:=
 \mathbf{R}_{\ \alpha \beta \tau }^{\tau}=
 \{R_{ij},R_{ia},\ R_{ai},\ R_{ab}\},$ and scalar curvature,
 $\ ^{s}\mathbf{R}:= \mathbf{g}^{\alpha \beta }\mathbf{R}_{\alpha \beta
}=R+S,$  for $R=g^{ij}R_{ij}$ and $S=h^{ab}R_{ab}$.
In Finsler like geometries, the well known Levi--Civita connection $\nabla,$ ($ \mathbf{T}[\nabla]=0$ and  $\nabla \mathbf{g}=0$), is not used  (it does not preserve the h-/v-- splitting).}

{\bf Theorem 1:}\  Any (pseudo) Finsler / Riemannian geometry, for a chosen N--connection $\mathbf{N}$,  can be modelled equivalently by geometric data $(\mathbf{g},\nabla)$ and/or $(\mathbf{N},\ ^F\mathbf{g},\ ^g\mathbf{D})$ if $\ ^g\mathbf{D}=\ ^g\nabla + \ ^g\mathbf{Z},$ when the connections and distortion tensor $\ ^g\mathbf{Z}$ are  determined  by $\mathbf{g}=\ ^F\mathbf{g}$.

{\it Example:} There is the Cartan (Finsler  type) d--connection ${\tilde{\mathbf{D}}}=(h\tilde{D},v\tilde{D})$   uniquely defined by the conditions $\tilde{\mathbf{D}}\ ^F\mathbf{g}=0$ and  $T_{\ jk}^{i}=T_{\ bc}^{a} =0$.\footnote{
 ${\tilde{\mathbf{D}}}$ is also a canonical almost symplectic connection used in deformation/ A--brane quantization (and two connection renormalization) of Einstein and generalized Finsler geometries \cite{vncbh}; its  N--adapted coefficients are computed ${\tilde \mathbf{\Gamma}}_{\gamma \tau}^\alpha =\{({\tilde L}_{jk}^{i}=\frac{1}{2}\ ^{F}g^{ir}(\mathbf{e}_{k}\
^{F}g_{jr}+\mathbf{e}_{j}\ ^{F}g_{kr}-\mathbf{e}_{r}\ ^{F}g_{jk}),\ {\tilde C}_{bc}^{a}=\frac{1}{2}\ ^{F}g^{ad}(e_{c}\ ^{F}g_{bd}+e_{c}\
^{F}g_{cd}-e_{d}\ ^{F}g_{bc})\}$.}
For instance, an EFG theory on $TV,$ or $\mathbf{V}$, can be constructed similarly to GR when ${\tilde{\mathbf{D}}}$ is taken instead of $\nabla$.

\subsection{Principles and axioms of EFG;\ fundamental field equations}
  A model of EFG with  $\mathbf{N}\sim \ ^F\mathbf{N}$ and metric compatible $\mathbf{D}$ uniquely defined by $\mathbf{g}\sim \ ^F\mathbf{g}$ (on $TV$, or $\mathbf{V}$) is derived similarly to GR and the standard particle model  following  such principles:
\begin{itemize}
  \item
      {\bf Generalized equivalence principle} for free fall and universality of gravitational redshift and related {\bf principle of general covariance} with "mixing" of Finsler  variables.
  \item
      {\bf Generalized Mach principle} for the  quantum energy/motion  encoded via  $(\mathbf{N},\mathbf{g},\mathbf{D})$ in aether spacetime models with anisotropic dependence on "velocity" type (nonholonomically constrained) variables $y^a$. This principle is less clear formulated both in GR and EFG.

  \item We can formulate a {\bf  constructive--axiomatic approximation} (as generalized Ehlers--Pirani --Schild, EPS, axioms)  with a generalized paradigm for "Lorentzian 4--manifold" in GR extended on  nonholonomic tangent bundles/ manifolds, for a corresponding model of EFG constructed in a form to be compatible with experimental data.
  \item Any {\bf motion equation and conservation laws} are derived from nonholonomc Bianchi identities for $\ \mathbf{D},$ with unique deformations $\nabla _i T^{ij}=0 \to \mathbf{D}_\alpha \Upsilon ^{\alpha \beta}\neq 0$, for generalized stress--energy tensors $\Upsilon ^{\alpha \beta}$ derived following a N--adapted variational/ covariant calculus.
  \item Generalized {\bf Einstein--Finsler/ Yang--Mills / Dirac} etc  equations  are formulated for certain data $(\mathbf{N},\mathbf{g},\mathbf{D})$. For zero distortions $\ ^g\mathbf{Z}$, we get standard constructions similar to those for  $(\mathbf{g},\nabla)$ in GR. The Einstein--Finsler (gravitational) field equations split as
\begin{equation}
R_{ij}-\frac{1}{2}(R+S)g_{ij} ={\Upsilon }_{ij},\
R_{ab}-\frac{1}{2}(R+S)h_{ab} ={\Upsilon }_{ab},\
R_{ai} ={\Upsilon }_{ai},\ R_{ia} = -{\Upsilon }_{ia}. \label{efeq}
\end{equation}%
      \end{itemize}
{\bf Remark:}  The  equations (\ref{efeq}) can  be integrated in very general forms and quantized, for instance, for the canonical/Cartan d--connection \cite{ijgmmp,vncbh,vprincipl}. Imposing  nonholonomic constraints, we can generate exact solutions in GR and/or quantum models with the Levi--Civita connection.

\section{Finsler Branes and Cosmological Solutions}
\label{s3}
{\it Diagonal brane configurations on $TV$} can be generated by ansatz
\begin{equation}
\mathbf{g} =\ \phi ^{2}(y^{5})\eta _{\alpha \beta }du^{\alpha }\otimes du^{\beta }-  \left( \mathit{l}_{P}\right) ^{2}\overline{h}(y^{5})[\ dy^{5}\otimes \
dy^{5}+dy^{6}\otimes \ dy^{6}\pm dy^{7}\otimes \ dy^{7}\pm dy^{8}\otimes \
dy^{8}], \label{ansdiag8d}
\end{equation}%
where $\eta _{\alpha \beta }=diag[\pm 1]$ and $\alpha ,\beta
,...=1,2,3,4$; with a "length constant" $\mathit{l}_{P}$.\footnote{There are used indices  $\ ^{1}\alpha =(\alpha ,5,6)$ and $\ ^{2}\alpha =(\ ^{1}\alpha ,7,8),$ respectively, for 6--d and 8--d models; indices  $\ ^{2}\alpha ,\ ^{2}\beta ,...$ run values $1,2,3,4,5,...,m$ where $m\geq 2$; $\ ^1a=5,6$ and $\ ^2a=7,8$;   coordinates $y^{5},y^{6},y^{7},y^{8}$ are velocity type (for instance, $y^{5} $ has a finite maximal value $y^{5}_0 $ on $TV$ because the light velocity is finite). The anholonomic deformation method allows us to construct solutions for any convenient signature; for instance, (non) homogeneous cosmological  metrics can be constructed for $u^2 =t$, or $u^3=t$.}
 Such a metric  is a solution of (\ref{efeq}) if
 $\phi ^{2}(y^{5})=\frac{3\epsilon ^{2}+a(y^{5})^{2}}{3\epsilon
^{2}+(y^{5})^{2}}$ and $\mathit{l}_{P}\sqrt{|\overline{h}(y^{5})|}=%
\frac{9\epsilon ^{4}}{\left[ 3\epsilon ^{2}+(y^{5})^{2}\right] ^{2}}$, where $a$ is an integration constant and the width of brane is
$\epsilon ^{2}=40M^{4}/3\Lambda $,
with some fixed integration parameters when $\frac{\partial ^{2}\phi }{%
\partial (y^{5})^{2}}\mid _{y^{5}=\epsilon }=0$ and $\mathit{l}_{P}\sqrt{|%
\overline{h}(y^{5})|}\mid _{y^{5}=0}=1;$ this states the conditions that on
diagonal branes the Minkowski metric  is 6--d or 8--d (for  $TV$). For a "bulk"
 cosmological constant $\Lambda $, the  nonzero
components of stress--energy tensor are chosen
 $\Upsilon _{\ \delta }^{\beta }=\Lambda -M^{-(m+2)}
\overline{K}_{1}(y^{5}),\Upsilon _{\ 5}^{5}=\Upsilon _{\
6}^{6}=\Lambda -M^{-(m+2)}\overline{K}_{2}(y^{5}),$
for a fundamental mass scale $M$ on $TV,$ $\dim TV=8$;  $\overline{K}_{1}$ and $\overline{K}_{2}$ must satisfy certain compatibility conditions with the field equations and  conservation laws  \cite{vprincipl}.

 We can construct {\it off--diagonal Finsler brane }
 trapping  solutions (with cosmological scenarios  for $h_{3}(x^i,y^3=v=t)$)   deforming the metric      (\ref{ansdiag8d})   to  ansatz with nontrivial
   N--connection coefficients:\
 $N_{i}^{3} = w_{i}(x^{k},v),N_{i}^{4}=n_{i}(x^{k},v)$;  $N_{\alpha}^{5} =\ ^{1}w_{\alpha}(x^{k},v,y^{5});$    $N_{\beta}^{6} = \ ^{1}n_{\beta}(x^{k},v,y^{5});$
   $N_{\ ^1\alpha}^{7} =\ ^{2}w_{\ ^1\alpha}(x^{k},v,y^{7})$; $N_{\ ^1\beta}^{8} = \ ^{2}n_{\ ^1\beta}(x^{k},v,y^{7})$.  Such EFG spaces are characterized  by metrics
\begin{equation}
\mathbf{g} = g_{i}dx^{i}\otimes dx^{i}+h_{a}%
\mathbf{e}^{a}{\otimes }\mathbf{e}^{a}+
 (\mathit{l}_{P}) ^{2} (\overline{h}/\phi ^{2}) [\
^{q}h_{\ ^1a}\ \mathbf{e}^{\ ^1a}\otimes \ \mathbf{e}^{\ ^1a}+
\ ^{q}h_{\ ^2a}\ \mathbf{e}^{\ ^2a}\otimes \ \mathbf{e}^{\ ^2a}] \label{odsol}
\end{equation}
where  $ \mathbf{e}^{3} =dy^{3}+w_{i}dx^{i},\mathbf{e}^{4}=dy^{4}+n_{i}dx^{i},
\ \mathbf{e}^{5} = dy^{5}+\ ^{1}w_{i}dx^{i},\
\mathbf{e}^{6}=dy^{6}+\ ^{1}n_{i}dx^{i},\
 \mathbf{e}^{7} = dy^{7}+\ ^{2}w_{i}dx^{i},\
 \mathbf{e}^{8}=dy^{8}+\ ^{2}n_{i}dx^{i}.$
 Explicit formulas for coefficients (see \cite{vprincipl}) are determined
  by "effective/polarized" cosmological constants
  induced by  sources in (\ref{efeq}),
 $\widetilde{\mathbf{\Upsilon }}_{\ \delta }^{\beta } =diag[\widetilde{%
\mathbf{\Upsilon }}_{\ 1}^{1}=\widetilde{\mathbf{\Upsilon }}_{\ 2}^{2}=%
\widetilde{\mathbf{\Upsilon }}_{2},\widetilde{\mathbf{%
\Upsilon }}_{\ 3}^{3}=\widetilde{\mathbf{\Upsilon }}_{\ 4}^{4}=\widetilde{%
\mathbf{\Upsilon }}_{4},\
 \widetilde{\mathbf{\Upsilon }}_{\ 5}^{5}=\widetilde{\mathbf{\Upsilon }}_{\
6}^{6}=\widetilde{\mathbf{\Upsilon }}_{6},\widetilde{%
\mathbf{\Upsilon }}_{\ 7}^{7}=\widetilde{\mathbf{\Upsilon }}_{\ 8}^{8}=%
\widetilde{\mathbf{\Upsilon }}_{8}]$ and for corresponding classes of generating and integration functions.

\section{Conclusions}
\label{s4}
The main goal of this work was to show how basic  principles and  axioms  for  the GR theory can be extended on nonholonomic tangent bundles/ manifolds. We provided a self--consistent scheme for formulating metric compatible Finsler gravity models in a form most closed to the standard theory of particle physics. Following the anholonomic deformation method, there were constructed new classes of  cosmological off--diagonal solutions  for Finsler branes (\ref{odsol}).

 We found that  metric compatible Finsler connections allows us to realize trapping gravitational configurations with  well--defined physical properties for a range of parameters. Such brane/trapping effects of  QG and/or LV depend on the mechanism of Finsler type gravitational and matter fields
interactions on tangent bundle $TV$ over a spacetime $V$ in GR  and it is expected that they may be detected in TeV physics, or via modifications in modern cosmology scenarios  and astrophysics.

\vskip3pt
\textbf{Acknowledgement:} I'm grateful to Organizers  ERE2010 and M. Sanchez Caja for supporting  my participation at Spanish Relativity Meeting, Granada (September 6-10, 2010).

\end{document}